# User Profile-Driven Data Warehouse Summary for Adaptive OLAP Queries

Rym Khemiri and Fadila Bentayeb

ERIC Laboratory, University of Lyon, Lumiere Lyon2
5 av. P. Mendes-France 69676 Bron Cedex
Lyon, France
`Rym.Khemiri@univ-lyon2.fr`
`Fadila.Bentayeb@univ-lyon2.fr`
`http://eric.univ-lyon2.fr`

**ABSTRACT**

*Data warehousing is an essential element of decision support systems. It aims at enabling the user knowledge to make better and faster daily business decisions. To improve this decision support system and to give more and more relevant information to the user, the need to integrate user's profiles into the data warehouse process becomes crucial. In this paper, we propose to exploit users' preferences as a basis for adapting OLAP (On-Line Analytical Processing) queries to the user. For this, we present a user profile-driven data warehouse approach that allows dening user's profile composed by his/her identifier and a set of his/her preferences. Our approach is based on a general data warehouse architecture and an adaptive OLAP analysis system. Our main idea consists in creating a data warehouse materialized view for each user with respect to his/her profile. This task is performed off-line when the user defines his/her profile for the first time. Then, when a user query is submitted to the data warehouse, the system deals with his/her data warehouse materialized view instead of the whole data warehouse. In other words, the data warehouse view summaries the data warehouse content for the user by taking into account his/her preferences. Moreover, we are implementing our data warehouse personalization approach under the SQL Server 2005 DBMS (DataBase Management System).*

**KEYWORDS**

*Adaptive OLAP queries, Content personalization, user profile, user preferences, data warehouse materialized view.*

## 1. INTRODUCTION

To provide users with only relevant data from the huge amount of available information, personalization systems resort to user preferences to allow users to express their interest on specific parts of data. Most often, user has two types of preferences: hard and soft preferences. Hard preferences are user constraints which must firstly be satisfied. Whereas, soft constraints represent optional constraints of the user i.e which can be satisfied in the second instance (after hard preferences). For instance, for a user who prefers black cars, he can insist on having a low-priced car (price < 10000 €). Therefore, the color represents a soft preference (optional constraint), however the price < 10000 € is an obligation (hard preference).

Assuming that the data warehouse is large and only a small part is of interest to the user, sorting the whole data warehouse for each query will result in both wasting resources and slow query responses. We argue that personalized data warehouse content aims to improve the process of querying by considering the user profile as an active part during the relevance evaluation process. Thus, in this paper, we address the problem of finding relevant data in the data warehouses based on user preferences. We propose preprocessing steps that can be used to reduce the online time for processing each query. In particular, given a set of user preferences, a data warehouse instance and a query, we are interested in providing the user with only the most preferable data in the data warehouse.

Our approach is in contrast with "one fits all" criterion. Nevertheless, as not all human preferences are similar and they usually change depends on situations. It is well known that user preferences are complicated, multiple, ever changing and they should be understood within a context which depends. Our main idea consists then in creating a data warehouse summary for each user with respect to his profile (set of preferences). The data warehouse summary aims at providing users with only the data that is of interest to them from the huge volume of available information in the data warehouse. Preferences have been used as a means to address this challenge. This task is performed offline when the user defines his/her profile for the first time. Then, when a user query is submitted to the data warehouse, the system deals with his data warehouse summary instead of the whole data warehouse. The data warehouse summary is a materialized view created w.r.t user preferences.

Our personalization approach is based on three steps: (1) defining users profiles as a set of preferences, (2) creating a data warehouse summary for each user profile and (3) exploiting this data warehouse summary during user queries process.

To validate our approach, we develop user profile-driven data warehouse personalization method inside Netbeans environment and using SQL Server 2005 as a DataBase Management System (DBMS). User profiles and their corresponding profile materialized views are stored into a profile database and a materialized views database respectively as a metadata.

The rest of the paper is organized as follows. Section 2 briefly reviews related works to personalization in some domains. Section 3 presents our approach and a motivating example. Then, Section 4 defines our personalization system architecture. In Section 5, we present our formalization. Our case study and experimental results are presented in Section 6. We propose some purposes of optimization in Section 7. Finally, we conclude the paper and present future directions in Section 8.

## 2. RELATED WORKS

Personalization has increasingly become an interesting topic in many domains. Its big challenge is to tailor queries, interfaces or contents to individual users' characteristics or preferences. The key idea behind personalization is to customize the query result according to specific user's interests. Preference query was introduced in database systems the first time in order to soften "the rigid way where the researched data characteristics must be specified" [25]. In the case where any object (any record) doesn't reply to these characteristics, it's nevertheless possible in some applications to accept objects having less good characteristics against research criteria.

After that, several extensive investigations were carried out and two major lines were be formed in literature for expressing preferences: quantitative and qualitative approaches [7]. In the qualitative approach, preferences are specified directly, whereas, in the quantitative approach, preferences are expressed indirectly by using scoring functions.

Since data warehouses are characterized by voluminous data and are based on a user-centered analysis process, including personalization into the data warehousing process becomes a new research issue [29]. Works in this domain are inspired from those proposed for personalization in Information Retreival (IR) [17, 12, 32, 37], Databases (DB) [8, 21, 6, 22], and Human computer Interaction (HCI) [16, 30]. Moreover, always inspired by web personalization, a personalizing keyword database search is proposed using user preferences [33]. Moreover, other personalized approaches resort to modelling user context in order to have contextual preferences [27, 34].

Current approaches for personalization on multidimensional databases focus on user definition with specific data as defined on traditional databases. For example, preferences for multidimensional data are defined as order relations [26] in the same way as defined in the context of databases [21] in order to define two methods of personalization before and after the query execution. This personalization approach is based on works related to OLAP query personalization using data filtering techniques based on user profile in order to refine queries by adding some predicates [3]. The objective of these works is is to take into account visualisation constraints to provide to the user a result focusing in his interest. Using annotations, a new personalization technique based on user preferences model is proposed in which weights are associated to multidimensional databases components [28]. The proposed conceptual model is based on multidimensional concepts (fact, dimension, hierarchy, measure, parameter or attribute). To assign priority weights to attributes of a multidimensional schema, the personalization rules are described using the Condition-Action formalism. More recently, this model has been used for handling the context notion in order to closely relating user requirements to their current context [18, 19].

In [4], author used data mining techniques as aggregation operators to update dimension hierarchies in data warehouses without taking into account user preferences.

Inspired by [21] and [14, 15] propose an approach to adapt preference constructors to multidimensional context. Formulated on schema, preferences can not only be expressed over attributes and thus over cuboids but also preferences can be expressed over numerical values (measures). The preferences composition is modelled using predicate logic attributes and expressed through Pareto composition (two preferences are equally relevant) or Prioritization (a preference is more relevant than another).

Garrigos et al. use the data warehouse multidimensional model, user model and rules for the data warehouse personalization [11]. As a result, a data warehouse user is able to work with a personalized OLAP schema, which matches his needs best of all. Based on ECA-rules (Event-Condition-Action) [35]), PRML (Personalization Rules Modeling Language, described in [10]) is used in [11] for specification of OLAP personalization rules. The structure of such PRML rule can be presented with following statement: when event do if condition then action endIf endWhen. After that, in [24], a new method was proposed which provides exhaustive description of interaction between user and data warehouse. A set of user-describing profiles (user preference, temporal, spatial, preferential and recommendational) have been developed. A metamodel which formulates user preferences for OLAP schema elements and aggregate functions has been proposed. This model reflects connections among user-describing profiles.

Recently, Aligon et al [2]( based on [14] and [15]) propose an approach which extract preferences from MDX query log by using datamining techniques. Thus, extracted preferences (in form of association rules) serve to annotate MDX queries in order to improve proactivness.

One may observe that in OLAP context, the most emerging axis is the query personalization. The main idea behind its process consists in considering the user preferences when answering query. Despite most research works in OLAP personalization have been focused on personalization while querying, there are few works which addressed content personalization before querying. Our approach is different, so that, we propose using preferences to personalize the content of the data warehouse where personalized content aims at improving the querying process by taking into account the particular interests of individual users. However, we will follow a qualitative approach, since we think it is more natural for the user to express preferences among predicates directly.

In this section we have reviewed the current approaches for personalization in data warehouses. We present a comparative table 1 confronting the panoply of the proposed approaches. We choose some criteria that we consider relevant to compare personalization approaches.

- *Source* : this criterion presents the object to exploit for personalization which can be a user profile (interests, preferences, constraints,...), query history (log file) or thes user context.

- *Personalization Time* : the time of personalization : before querying, while querying or after querying.

- *Personalization Object* : this criterion presents the object of the proposed method if it is a query, an interface or a content to personalize.

- *Input Data* : this criterion presents the inputs of the proposed method if it is DW schema or DW instance or both of them.

- *Output Data* : this criterion presents the outputs of the proposed method if it is a query, a set of tuples or a personalized schema.

Table 1. Survey of OLAP personalization approaches

| Research Works | | Bellatreche et al. 2005 | Bentayeb 2008 | Jerbi et al. 2008, 2009 Ravat and Teste 2008 | Garrigos et al. 2007 | Kozmina et al. 2010 | Golfarelli et al. 2009, 2011 | Aligon et al. 2011 | Our approach |
|---|---|---|---|---|---|---|---|---|---|
| **Source** | User profile | × | | × | × | | × | | × |
| | Query Log | | | | | × | | × | |
| | Context | | | × | | | | | |
| **Time (% querying)** | Before | × | × | × | | × | | × | × |
| | While | | | | × | × | | | |
| | After | | × | | | | | × | |
| **Object** | Query | × | × | × | × | × | | × | × |
| | Interface | × | | | | | | | |
| | Content | | | | | | | | × |
| **Input Data** | DW schema | | × | × | × | × | | | |
| | DW instance | × | | | | | × | × | × |
| **Output Data** | Query | × | | × | | | × | | |
| | Tuples | | | | | | × | | × |
| | Schema | | × | | × | | | | |

## 3. USER PROFILE-DRIVEN DATA WAREHOUSE SUMMARY

In our approach, we focus on content personalization in order to provide the relevant data to the decision maker according to his profile. The personalized content is the content which matches a particular profile. It takes into account preferences contained in this profile to correctly generate data warehouse summary. This later is in the form of a materialized view retrieving the most appropriate content to the user preferences. In this section, we firstly define the user profile including his/her preferences. Secondly, we present the general principle of our approach which we motivate it by an example.

### 3.1 User profile

To personalize data warehouses, it is essential to obtain knowledge about users in order to narrow their exploration space. These data characterizing a user are often organized and structured in a model called user model or user profile. These users profiles are vital in adaptive systems that personalize the human computer interaction.

Preferences are usually considered as expressions which enable to order elements of the information [5, 6, 36, 20] and most often, user profile is reduced to user preferences. Then, a user profile is a set of descriptors including which a user envisage to fulfil in the system, how to do it, what is the type and the order of the obtained results and how they can be displayed.

The qualitative approach was bound to specify directly preferences between tuples in the response to the query, typically by using preference binary relations. An example of preference relation is : "prefer a tuple of product to another if they have the same reference but the price of the first is cheaper than the second". In the other hand, the qualitative approach expresses preferences by using score functions which associate a numeric score to every tuple of the query result. Thus, the tuple $t_1$ is preferred to the tuple $t_2$ if and only if the score of $t_1$ is upper than the score of $t_2$.

In literature, problems discussed for preferences show duality expressiveness/efficiency. In fact, the major difficulty in using preferences is their expressiveness, especially when they are quantitative. The other practical difficulty is algorithmic efficiency of preferences computing. Therefore, to avoid these difficulties, many works have tried to exploit user profile. Some people used languages and preference operators such as the interesting SQL extension, called PRFERENCE SQL [21] or the preference operators such as SKYLINE [5] and winnow [7]. In addition, some people resorted to query rewriting or merely query enhancement[22, 23] which consists in integrating into the user query some elements from the user profile. This technique is well used in Information Retrieval domain [31] and this is very recent in database domain.

A data is relevant if and only if it fulfils at least user preferences. Sometimes, to be relevant, a data must respect user context. Although, user context represents a big problem: How to represent it? In [1], "Context can be represented and processed" but for Dourish [9] context is not representable.

## 3.2 General principle
A data warehouse often contains vast amounts of information that is difficult for a new user to comprehend. It is necessary to employ a personalization mechanism to assist users. What part of data warehouse should be provided initially for a new user without any DW experience ? A new user can express his/her preferences. Then a summary of the data warehouse can be proposed to the user based on his/her preferences. For experienced users, what information can be provided to turn them into frequent users ? Our proposed personalization mechanism provide a data warehouse summary that encourage further utilization of the data warehouse.

The originality of our approach consists in creating a materialized view for each user profile. The obtained profile-driven materialized view contains only the part of the data warehouse which fulfils user hard preferences. In this case, when a user OLAP query is submitted on the data warehouse, the system deals with the materialized view instead of the whole data warehouse.

Our approach presents many advantages. It offers the possibility for each user to define his/her specific preferences on the data warehouse, to get quick answers of his/her OLAP queries by using profile-driven data warehouse materialized views.

Our approach is based on three steps: (1) defining users profiles as a set of preferences, (2) for each user profile, a data warehouse materialized view is created and (3) exploiting this data warehouse materialized view during the user queries process.

## 3.3 Motivating example
In this section, we illustrate the basic idea of our approach with a simple example of Sales Data Warehouse of car advertisements (1). In fact, nowadays, buying a new or used car can be a challenge, but there's plenty of help to make the decision of buying the right car. In fact, if a user want to buy a used or new car from a private owner or a dealer, there is many sites on Internet

International Journal of Database Management Systems ( IJDMS ) Vol.4, No.6, December 2012

which are full of ads. A car advertisement might have metadata parameters (key things to look for in the ads) that describe its price, mileage, model year, color, date of the last inspection, date of first registration, number of owners, energy, fuel, etc.Therefore, we should help the user to have the appropriate decision.

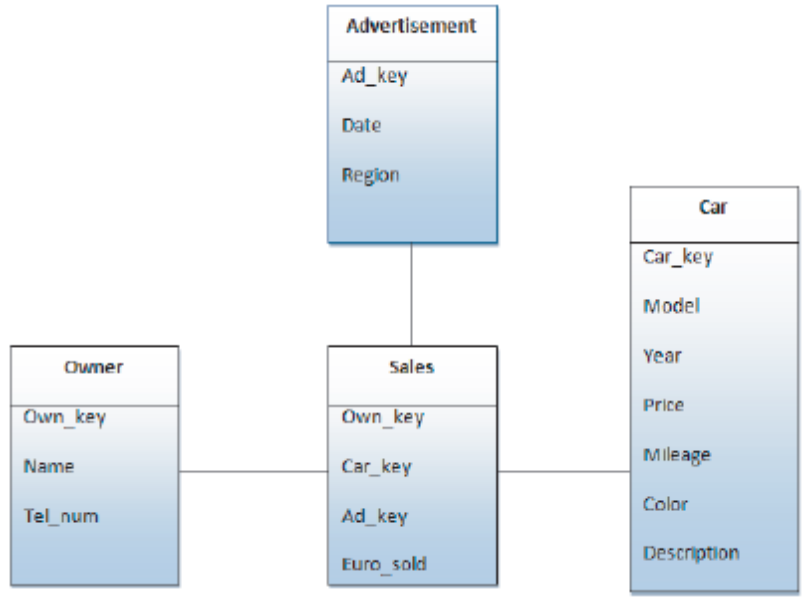

Fig.1. Sales Data Warehouse Star Schema

In order to satisfy the user, firstly we must satisfy his/her hard preferences (essential characteristics) and after that he can select the appropriate ones among the returned results by making comparison between his/her soft preferences (optional characteristics). Figure 1 depicts our sales data warehouse model. It contains three dimensions one fact table 'Sales' that includes, besides foreign keys, one measure 'euro-sold' which represents the sales earnings. Our dimensions collection includes:

- Car Dimension: it represents all car features (model, year, price, etc...).
- Owner Dimension: it represents the owner of the car.
- Advertisement Dimension: it defines the ad by its date and its region according to the owner address.

To choose a car, the system must identify the user preferences. It will narrow the exploration space of the user. To buy a car, for a user the model Mercedes and diesel energy are essentials. For another user, he can be interested in cars having as model BMW, price < 20000 €, black color and the advertisement region is 'Rhone'. Thus, user must be aware of handling his/her preferences on the data warehouse especially his/her hard ones. When he/she connects to the system, he/she must enter his/her preferences especially his/her hard ones.

For example, let a user who insists on finding a car which its model year > 2007, its price < 20000 € and has black color. User hard preferences are his constraints which can be represented in predicates on dimensions: For Car dimension, user preferences are defined by predicates as follows:

- Car.year > 2007
- Car.price < 20000
- Car.color = 'black'

For Advertisement dimension, user preferences are defined by predicates as follows:

- Advertisement.region='Rhone-Alpes'
- For dimension Owner, user haven't any requirements.

To identify a list of cars that meet user criteria, our system will submit queries for searching cars which fit more user constraints. The user will have a list of cars (satisfying his constraints) from which to start. In fact, when the user connects to the data warehouse the first time, he should enter his/her preferences. Straight afterwards, the system, on off-line mode, submits queries including in condition these preferences already entered by the user in the form of predicates. For our example, based on user preferences, the system will submit these queries:
For Car dimension:

Query:
Select *
From Car                    Car dimension
Where $Car.year > 2007$      First Preference
and   $Car.price < 20000$    Second Preference
and $Car.color =' black'$;   Third Preference

For Advertisement dimension, the system will submit this query:

Query:
Select *
From Advertisement          Advertisement Dimension
Where Advertisement='Rhones-Alpes';  Preference

As a result, the system will have for each dimension only data fulfilling the user preferences (conditions of the two queries) on this dimension. Our personalization system will not take these dimensions data apart, it will join them in order to have the data warehouse materialized view for this user.

In on-line mode, if the user inputs a query, but a very broad one. For example, a wide query concept such as "Select * from car", the system will handle only the data warehouse materialized view already created for this user instead of the whole data warehouse. Although, if the user submits a narrower query like "select * from car where model='BMW'". This condition does not exist in user hard preferences already entered by the user himself. We can consider this condition as a soft preference since it presents a user option. So, the system will directly handle the data warehouse materialized view of this user and cut down the set of results by this condition: Car.model='BMW'. Finally, the system will provide answer to the user query. As a consequence, if the user make a wide query, he gives more room for personalization than a narrower one.

International Journal of Database Management Systems ( IJDMS ) Vol.4, No.6, December 2012

## 4. PROFILE-DRIVEN DATA WAREHOUSE ARCHITECTURE

In our study, a personalized approach based on user profile is proposed to provide adaptive answers to the data warehouse queries. Figure 4 depicts our system architecture which can be divided into two main parts according to system operation procedures i.e front-end and back-end parts. Front-end part manages communications with decision makers and records their data. On the other hand, back-end part aims to analyse decision makers characteristics and to provide their profiles according to data entered by them on the front-end part. The user interface belong to the front-end part. It can identify user preferences and the corresponding data warehouse materialized view, transfers users' queries to system and return the suggested results. It provides a friendly human-machine interaction. As a consequence, the data warehouse materialized view is generated on off-line mode. One of the most important parts of data warehouse are its metadata. Metadata, which are used by developers in order to manage and control the creation and maintenance of the data warehouse, are kept outside of the data warehouse. The metadata concerning data warehouse users are on the contrary a part of data warehouse. This data are used to control access to and analysis of data. Therefore, metadata, besides describing contents of the data warehouse, it can provide the user with information useful in personalizing the content. Consequently, we have two metadata : user profile database (UP DB) and user profile materialized view database (UPV DB).

In order to identify decision makers, the user should make a user authentication (UA). If he is a beginner i.e the first time he connects to the system so he can enter his preferences and the system records his profile (set of preferences) in the user profile database (UP DB) i.e all information about the user is stored in user profile database. If the data warehouse materialized view is already created, it will be automatically stored in user profile materialized view database.

Although, if after authentication, the system identify the user as a experienced i.e it's not his first connection or interaction with the system, the system will deal directly with his/her materialized data warehouse materialized view stored in the user profile materialized view database (UPV DB).

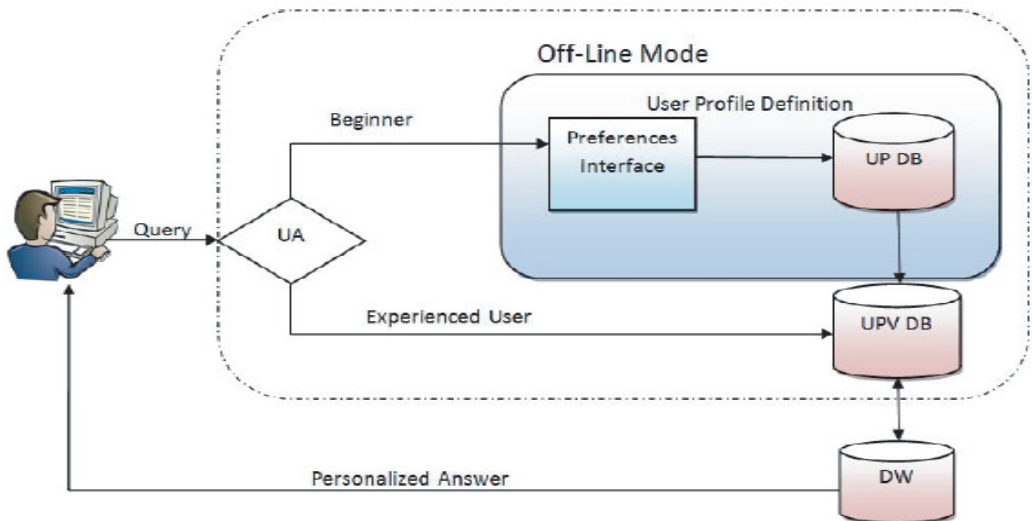

Fig. 2. User profile-driven data warehouse materialized view architecture

### 4.1 User profile-driven data warehouse materialized view creation

Our goal is to create for each user his/her own data warehouse materialized view based in his/her profile. On off-line mode, our system can provide the data warehouse materialized view for a specific user. As long as the data warehouse represent a large amount of data, this personal data warehouse materialized view will narrow exploration space of the user. This part is realized when the user is a beginner or if he modifies his preferences.

Moreover, our personalization process can be depicted in figure 4. First, our personalization system must identify the user (UA). After the user enters his/her preferences, the system will estimate his/her interests, generate the corresponding data warehouse materialized view and store it in the user profile database (UP DB). If the user make a query, the system will use this materialized view else if the user logout the system, the materialized data warehouse view will be introduced in the next session.

In other words, if the user is a beginner, he must enter his preferences. As a consequence the system will create his profile and create his personal materialized view according to the profile. Else, if the user is experienced so the system will directly use his/her personal materialized view.

### 4.2 User profile-driven data warehouse materialized view exploitation

User can exploit his/her personalized data warehouse summary while querying task. He can submit OLAP queries with regard to our personalization approach. The user profile materialized views aim at impacting the OLAP queries results improving displayed data according to users preferences. For example, if we have many colours of cars, and we have determined that the user prefers black color, we should probably want to display black cars to the user (assuming we want to have happy users).

In our case, the context to match against includes at least the user's profile, the user's current request and the user's current session. If it's his/her first connection, the system will create a new profile initialized with any values and ask the user to complete it. straight afterwards, the system will store this profile in the a database (UP DB). Then, the system will create a data warehouse materialized view for each user with respect to his profile. Moreover, the process that is responsible for the user materialized view creation is computed in the off-line part of the application. This means that its algorithms are not optimized to run in real- time. More specifically, the profile creation module receives as input user preferences and produces as output the corresponding user materialized view. If the user authenticates (logs in), the system will put the profile in the session. The profile is kept in the session. If the user makes a request, the system deals with his/her data warehouse summary (data warehouse materialized view) instead of the whole data warehouse.

## 5. FORMALIZATION

To define user profile, we use a finite set of user preferences. We distinguish between two types of preferences: soft and hard ones. A soft preference is a current or temporary preference, while a hard preference consists of permanent constraint. The result of each query depends on the user submitting it. Each user is modelled by his/her identity and his/her preferences. Thus, our user profile is the couple (user identity, user hard preferences). Hard preferences are specified by the user himself, whereas soft preferences can be identified in the query submitted by the user. To specify preferences, we use a simple preference model composed by predicates followed by operators and values. In particular, users express their preference for sets of tuples specified using selection conditions on some of the attributes of the tuples by assigning to a value using an operator. This operator is an element including this set $\{=,<,<,\leq,\geq,\neq\}$ in order to express interest for the specified tuples.

Let a data warehouse schema DW ($F, D_1, D_2, \ldots D_n$) where F is the fact table and D is dimension having as schema D ($A_1, A_2, \ldots, A_d$), where A is a dimension attribute.

**Definition 1.** User Preference on Dimension. Given a dimension schema D (A1,A2,…Ad), a preference p on Dimension D is a triple (Pred, operartor, value), where Pred is a predicate of the form Dimension.attribute ($D_i.A_{ij}$) that specifies conditions by operators operator on the values $a_{i_j} \in dom(A_{i_j})$ of attributes $A_{i_j}, 1 < i_j < d,$ of Dimension D.

The meaning of such a dimension user preference is that all dimension tuples t for which Pred holds are assigned the indicated value by the indicated operator. In our motivating example, we assume operator $\in \{ =, <, <, \leq, \geq, \neq \}$ for the Atrributes $A_{ij}$. For instance the preference (car.color, = , all) means that the user likes all colours of car.

There may be more than one preference applicable to a specific dimension D. In other words, a tuple t may satisfy the predicate part Pred of more than one of the preferences specified for a dimension D. In general, if more than one preference is applicable to a dimension, we choose the tuple which satisfy all the preferences or satisfying the majority.

If none of preferences are specified by the user to be be applicable to dimension tuples, then all dimension tuples are assigned to the materialized view. This is because, we consider preferences expressed by users to be indicators of positive interest and hard constraints. Consequently, we assume that an unexpressed preference on a given dimension, means that user has interest on the whole dimension. If we have preferences on a dimension, so we will cut down the number of tuples of this dimension. However, if we haven't any preferences on a dimension, we will take all its tuples;

We call the set of dimensions preferences that holds for a data warehouse, profile P.We assume that such profiles are available. In practice, preferences may be, for example, given by the users explicitly. The practical way to create P, is by giving a default profile to the user and allow him to update it appropriately.

Now, given user profile, we would like to use user preferences in the profile P in order to compute a data warehouse materialized view for the user based on these preferences. This problem is complicated by the fact that we have a set of dimensions and preferences on these dimensions.

Next, we define the Dimension Vector with regards to a set of user preferences on this dimension:

**Definition 2.** Dimension Vector. Let $p_{i_j}$ be a user preference on dimension $D_i$, a dimension vector $Ve_{D_i}$ is a set of tuples $t_i \in D_i$ such that $p_i[t_i]$ holds.

**Definition 3.** DW Materialized View. A data warehouse materialized view is defined as follows:
$V_{Dw} = \cap(Ve_{D_1}, Ve_{D_2}, ..., Ve_{D_m})$ where $Ve_{D_i}$ Dimension Vector and $D_i$ dimensions of our model.
Else, data warehouse materialized view can be defined as follows:
$V_{Dw} = \bowtie (Ve_{D_1}, Ve_{D_2}, ..., Ve_{D_m})$ where $Ve_{D_i}$ Dimension Vector and $D_i$ dimensions of our model.

Assume a data warehouse instance DW, a profile P and a query q. The DW materialized view creation problem is to rank all tuples t in the result of q with respect to user preferences. For computing all tuples in the result set, a solution that involves no pre-computation is to first find the set tuples respecting hard preferences, compute this set to have result of the query. Performance can be improved by performing preprocessing steps off-line.

## 6. IMPLEMENTATION

The hardware requirements for the realisation of our data warehouse personalization approach do not deviate from the hardware requirements from an usual data warehouse system. Since this software-based personalization model was developed using 'JAVA' on Netbeans environment on top of SQL Server 2005 DBMS. We selected this platform because it is independent, meaning that the current version runs on Windows and UNIX. In addition,we are using the Foodmart data warehouse as example, its schema is depicted by the Figure 3. Sales Fact 1998 is the main fact table that has sales information by store/location, product, time, customer, and promotion. Correspondingly there are 5 dimension tables joined to the fact table through foreign keys in the star schema.

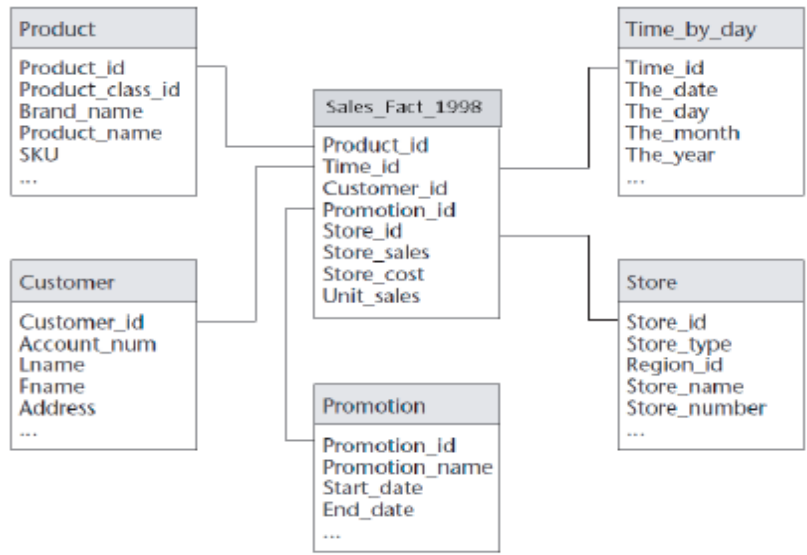

Fig. 3. Excerpt of FoodMart data warehouse

Our personalization system is based on several components:

- The user interface which allows users to interact with UP DB. Users define their preferences on data warehouse dimensions. Figure 4 depicts a screenshot from this interface.
- The UP database which contains user preferences
- The UPV database which contains materialized views generated by the system in off-line mode.

In order to identify decision makers, the user should go through authentication(UA). If he is a beginner i.e the first time he connects to the system so he can enter his/her preferences. Straight afterwards, the system records his/her profile (set of preferences) in the user profile database (UP DB). Then, the system creates the user-specific data warehouse view based on his/her profile and stores it in the UPV DB.

As explained in the introductory sections, automatic personalization is not appropriate in all situations. Therefore it is considered an optional feature that users can turn on and of at any time. As a matter of fact, in the current system, a slider in the main window allows the user to set the degree of personalization. Despite its advantages, data warehouse personalization if it exceeds we will risk to have over personalization. We plan to have the personalization process as an option to be turned by the user himself. Figure 4 shows that the personalized search represent an option. If it is turned on, the system creates the materialized view according to the selected user preferences.

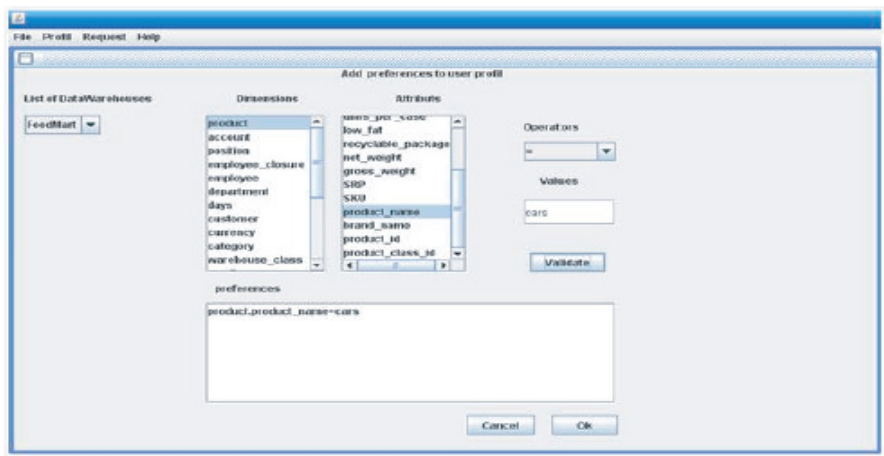

Fig. 4. User Preferences Interface

## 7. OPTIMIZATION

The main goal in data warehousing applications is optimizing the system performances [13]. Actually, the number of materialized views to build and stored (one per user) is the first drawback of our personalization method. Although, on one hand our materialized view represent only relevant data and on the other hand the number of users for a data warehouse is often limited. As a consequence this drawback does not arise a big problem and if it does, our approach supports help for decision which can classify users into various groups; these groups support the same information (group profile). For example, we could define a user group which has common preferences. In some cases, when personalizing content, we can use these groups.

Computing all data warehouse tuples for each query will result in both wasting resources and slow query responses. Since, user has hard constraints which must be applied on all queries and the number of tuples satisfying user constraints can still be large, we could instead compute the part which respects these constraints stored in the profile, we can extract identifiers of tuples in order to precompute materialized view. The approach constructs identifiers of tuples respecting user preferences. After constructing the user materialized view, we apply the submitted query in this part. Furthermore, instead of storing all tuples for each user, we just store the identifiers of these tuples. Then, for each query, we can search in this materialized view and quickly provide the appropriate results.

Moreover, to have the data warehouse materialized view, the dimension vectors are not joined to each other. In contrast, the data warehouse materialized view is created due to a star query. In fact, the star query is a join between the fact table Sales and the surrounding collection of dimension tables Car, Advertisement and Owner. Therefore, each dimension table is joined to the fact table Sales using a primary key to foreign key join.

## 8. CONCLUSION AND PERSPECTIVES

Commonly used to enhance user satisfaction, personalization is a means of meeting the user's needs more effiectively and efficiently, making interactions faster and easier and, consequently, increasing user satisfaction and the likelihood of user satisfaction.

In this paper, we have presented a data warehouse personalized approach because we believe that data warehouse personalization can bring us the user satisfaction and so the system success. This paper focused on adaptive data warehouse which helps to provide users with meaningful content when making decisions.

In the second part of this paper, a personalization system was introduced. This prototype implements a personalization model based on metadata. This models assigns one materialized view to each user. The overall goal of the personalization framework described in this paper is to represent, capture, and manage any information about data warehouse that can be exploited to improve the overall system services in different ways. The identification of the opportunities for improvement by personalization, and their achievement or facilitation, are part of the goals of this research. Actually, the contribution of this work is twofold: Data warehouse materialized view construction and the exploitation of this materialized view. In general, if more than one preference is applicable to a dimension, we choose the tuple which satisfy all the preferences or satisfying the majority. As a tool, we can resort to skyline operator.

Finally, let us point out that as such models mature, there are additional avenues of research that need to be pursued. To provide users with only relevant data from the huge amount of available information, personalization systems use preferences to allow users to express their interest on specific data. Most often, user preferences vary depending on the circumstances. For instance, decision makers requirements in winter may be different from those in summer. Therefore, user profile change over time with user context changing. As a consequence, currently, we think of creation of user profile versions which can address this challenge by supporting preferences that depend on user context such as the surrounding environment, time or location. These profiles versions, in contrast to long-term user profiles, represent contextual user profiles that capture user context when he is making his request. In our approach the user become aware of handling his/her preferences on the data warehouse. Problem will be on updating new content available or if the user is not aware of noting his/her preferences in some cases.